\documentclass[12pt]{article}
\usepackage{amsmath}
\usepackage{color}
\usepackage{fancyhdr}
\usepackage{verbatim}
\usepackage{amsfonts,longtable}
\usepackage{epsfig}
\usepackage{amsfonts}
\usepackage{color}
\usepackage[numbers,sort&compress]{natbib}

\def\R{\hbox{{\rm I}\kern-0.2em{\rm R}\kern0.2em}}

\def\bn{\begin{equation}}
\def\en{\end{equation}}
\def\bny{\begin{eqnarray}}
\def\eny{\end{eqnarray}}
\def\be{\begin{eqnarray*}}
\def\ee{\end{eqnarray*}}
\def\bc{\begin{center}}
\def\ec{\end{center}}

\def\({\left(}
\def\){\right  )}
\def\[{\left[}
\def\]{\right]}
\def\bc{\begin{center}}
\def\ec{\end{center}}

\newtheorem{dfn}{Definition}[section]
\newtheorem{thm}{Theorem}[section]
\newtheorem{rem}{Remark}[section]
\newtheorem{pro}{Proposition}[section]

\newtheorem{cor}{Corollary}[section]
\newtheorem{lem}{Lemma}[section]
\newtheorem{exm}{Example}[section]

\def\bn{\begin{equation}}
\def\en{\end{equation}}
\def\bny{\begin{eqnarray}}
\def\eny{\end{eqnarray}}
\def\be{\begin{eqnarray*}}
\def\ee{\end{eqnarray*}}
\def\bdn{\begin{dfn}}
\def\edn{\end{dfn}}
\def\btm{\begin{thm}}
\def\etm{\end{thm}}
\def\bpf{\begin{proof}}
\def\epf{\end{proof}}
\def\bpn{\begin{pro}}
\def\epn{\end{pro}}
\def\brk{\begin{rem}}
\def\erk{\end{rem}}
\def\bcy{\begin{cor}}
\def\ecy{\end{cor}}
\def\blm{\begin{lem}}\def\elm{\end{lem}}
\def\bex{\begin{exm}}
\def\eex{\end{exm}}

 \def\R{{\hat R}}
\begin{document}

\bc {\bf Symmetries and Reduction of Discrete Painlev\'e Equations
  }\ec
\medskip
\bc
M. Folly-Gbetoula$^{1,}$\footnote{Mensah.Folly-Gbetoula@wits.ac.za} and  A. H. Kara$^{1,2,}$\footnote{Abdul.Kara@wits.ac.za }
\\$^1$School of Mathematics, University of the Witwatersrand, Johannesburg, South Africa.\\
$^2$ Department of Mathematics and Statistics, King Fahd University of Petroleum and Minerals, Dhahran, Saudi Arabia.
\ec
\begin{abstract}
The construction and role of symmetries  for difference equations are now well known.
In this paper, the symmetry analysis of the discrete Painlev\'e equations is considered. We assume that the characteristics depend on $n$ and $u_n$ only and we obtain a number of symmetries. These symmetries are used to construct exact solutions in some cases.
\end{abstract}
\section{Introduction} \setcounter{equation}{0}
\noindent The application of Lie group theory to differential equations became a center of interest to many authors since the preliminary works of Sophie Lie. 
  Recently, a lot of work has been done 
  to extend the concepts to difference equations and some interesting results can be found in \cite{Hydon1,FK2,s}. 
It is now known that the association of  conservation laws, integrability and symmetries for difference equations is as important and conclusive as was established for differential equations even in the non-variational case. Symmetries are useful tools in the sense that they can be used to reduce the order of the equations via the difference invariance. Once the solutions of the reduced equations are obtained, one can readily recover the solutions of the initial equations using the reciprocal bijection. On the other hand, the method of discretization of differential equations is now well documented. Ramani in his paper \cite{Ramani} presented clearly his method for obtaining the discrete versions of the Painlev\'e equations and showed that the mappings for the discrete versions satisfy the same reduction relations as in the case of the continuous ones. In this paper, we construct symmetries of  these discrete versions of  Painlev\'e  equations and in some cases the symmetries are used to find exact solutions to the equations.
\section{Preliminaries and Definitions}
In this section we aim to provide  background on Lie symmetries analysis and  the algorithm that generates symmetries of ordinary difference equations. The operator $\mathcal{S}^{(i)}$ defined by
\begin{equation}\label{}
\mathcal{S}^{(i)}:n \mapsto n+i
\end{equation}
will be refereed to as the shift operator.\par
\noindent Consider a $p$th-order difference equation
\begin{equation}\label{E}
u_{n+p}=\omega(n, u_n, u_{n+1}, \dots, u_{n+p-1})
\end{equation}
and the transformation
\begin{equation}\label{}
\Gamma:(n, u_n, u_{n+1}, \dots, u_{n+p})\mapsto(\hat{n}, \hat{u_n}, \hat{u}_{n+1}, \dots, \hat{u}_{n+p}).
\end{equation}
We search for a one-parameter Lie group of transformations of the form
\begin{eqnarray}
\hat{n}&=&n+\epsilon \, \xi(n, u_n, u_{n+1}, \dots, u_{n+p}),\\
\hat{u}_{n+i}&=&u_{n+i}+\epsilon \, \mathcal{S}^{(i)}Q(n, u_n, u_{n+1}, \dots, u_{n+p})\label{transfo1}
\end{eqnarray}
for $0\leq i\leq p$. $Q$ is the characteristic and $\epsilon$ the parameter of the group transformation $\Gamma$. To ease our computation we shall assume that the characteristic depends on $n$ and $u_n$ only. It is worthwhile to mention that not all equations admit such characteristics. Then, the linearized symmetry condition
\begin{eqnarray}\label{symcdt}
\mathcal{S}^{(p)} Q= X \omega
\end{eqnarray}
is obtained by expanding the transformed equation
\begin{equation}\label{E'}
\hat{u}_{n+p}=\omega(\hat{n}, \hat{u}_n, \hat{u}_{n+1}, \dots, \hat{u}_{n+p-1})
\end{equation}
to first-order in $\epsilon$. Here, the corresponding generator will be
\begin{eqnarray}\label{gener}
X= \xi(n,u_n)\partial n + Q(n,u_u)\partial u_n.
\end{eqnarray}
In order to solve equation (\ref{E}), we require the $p$th extension of $X$ and we further impose the symmetry condition (\ref{symcdt}) whenever (\ref{E}) holds. Usually, this leads to an equation that involves functions with different arguments which makes the obtention of solutions difficult. However, the solutions can be obtained after a series of lengthy calculations. Despite the fact that  painlev\'e equations are second-order difference equations,  the procedure for obtaining symmetries is still lengthy and some details of computations will not be presented in this paper. The main steps involved are three; firstly differentiate (\ref{E}) implicitly with respect to $u_{n}$ keeping $\omega$ fixed; secondly differentiate with respect to $u_n$ ( keeping $u_{n+1}$ fixed) as many as possible in order to get rid of any undesirable arguments; thirdly we equate all coefficients of powers of $u_{n+1}$ to zero since the characteristic $Q$ does not depend on $u_{n+1}$. These steps normally lead to a system of determining equations whose solutions give the expressions of $\xi$ and $Q$. The number of independent unknown constants determines the dimension of the group. Once the symmetries are found we use the method of characteristics given by
\begin{equation}\label{charac}
\frac{du_n}{Q} = \frac{du_{n+1}}{\mathcal{S}Q}=\frac{du_{n+2}}{\mathcal{S}^{(2)}Q}=\dots =\frac{du_{n+p-1}}{\mathcal{S}^{(p-1)}Q}
\end{equation}
to obtain the invariant functions which are actually the first integrals of (\ref{charac}).

\section{Symmetries of the Discrete Painlev\'e Equations}
In this section we consider the discrete forms of the Painlev\'e equations and we aim to find their symmetries using the method explained above. As mentioned earlier, we shall assume that the characteristics $\xi$ and $Q$ depend on $n$ and $u_n$. In some cases the exact solutions are found.
\subsection{The discrete Painlev\'e I}
The discrete form  is known to be
\begin{equation}\label{dp1'}
u_{n+2}=\omega =- u_{n}-u_{n+1} +\frac{an+b}{u_{n+1}}+c
\end{equation}
for some constants $a,\,b$ and $c$ (we shall assume that $c$ is zero). We assume that the symmetry generator is of the form $X=\xi(n,u_n)\partial n+Q(n,u_n)\partial{u_n}$. Imposing the symmetry condition (\ref{symcdt}) we have
\begin{eqnarray}\label{scp1b}
Q(n+2,\omega)+Q(n+1, u_{n+1})\left(1+\frac{an+b}{u_{n+1}^2}\right)+Q+\frac{a}{u_{n+1}} \xi&=& 0.
\end{eqnarray}
To solve for $Q$ and $\xi$ we first differentiate equation (\ref{scp1b}) implicitly with respect to $u_n$ to get
\begin{eqnarray}\label{scp1c}
&&Q'(n,u_n)-\xi'(u,u_n)\frac{a}{u_{n+1}}-\frac{{u_{n+1}}^2}{{u_{n+1}}^2+an+b}\bigg[-\frac{2(an+b)}{{u_{n+1}}^3}Q(n+1,u_{n+1})\nonumber\\ &&+\left(1+\frac{an+b}{u_{n+1}^2} \right)Q'(n+1,u_{n+1})+\frac{a}{{u_{n+1}}^2} \xi(n,u_n)\bigg]+= 0.
\end{eqnarray}
Secondly, we differentiate (\ref{scp1c}) with respect to $u_n$ keeping $u_{n+1}$ fixed. This leads to a second-order differential equation that involves two different functions with same arguments $n$ and $u_n$:
\begin{eqnarray}\label{scp1d}
&&{u_{n+1}}^2Q''+(an+b)Q''-\left(au_{n+1}+\frac{an^2+ab}{u_{n+1}}\right)\xi''
-a\xi'= 0.
\end{eqnarray}
In the equation above, $Q$ and $\xi$ do not depend on $u_{n+1}$, therefore we can equate all the coefficients of all the powers of $u_{n+1}$ to zero. We get
\begin{eqnarray}
{u_{k+1}}^{-1}&:&-a(an+b)\xi'' =0\\
1&:& (an+b)Q''-a\xi' =0\\
u_{k+1}&:&-a\xi''=0\label{deteqdp1}\\
{u_{k+1}}^{2}&:&Q'' =0.\label{deteqdp1'}
\end{eqnarray}
\subsubsection{Case $a \neq  0$}
We have
\begin{eqnarray}\label{scp1e'}
Q= \alpha u_n +\beta, \qquad \qquad \xi = \gamma ,
\end{eqnarray}
where $\alpha, \beta$ and $\gamma$ are functions of $n$. We substitute (\ref{scp1e'}) back in equations (\ref{scp1d}), (\ref{scp1c}) and (\ref{scp1b}). We obtain that
\begin{eqnarray}\label{scp1f}
\alpha= A, \qquad \beta=0, \qquad \gamma=\frac{2\alpha}{a}(an+b).
\end{eqnarray}
Therefore the symmetry generator of (\ref{dp1'}) is given by
\begin{equation}
X=2(an+b)\partial n+au_n\partial{u_n}+au_{n+1}\partial{u_{n+1}}.
\end{equation}
\subsubsection{Case $a=b=0$}
Equation (\ref{dp1'}) becomes
\begin{eqnarray}\label{dp1=0}
u_{n+2}=-u_n-u_{n+1}
\end{eqnarray}
and solutions to equations (\ref{deteqdp1}) and (\ref{deteqdp1'}) in this case are given by
\begin{eqnarray}\label{scp1e}
Q= c_1 u_n +\theta, \qquad \qquad \xi = c_2 ,
\end{eqnarray}
where $c_1,\, c_2$ are constant and $\theta$ is a function of $n$  that satisfies the original equation. 
Here, we obtain four symmetries given as follows
\begin{eqnarray}
X_1&=&u_n \frac{\partial}{\partial u_n} + u_{n+1} \frac{\partial}{\partial u_{n+1}},\\
X_2&=&(-1)^n\left[\cos\left(\frac{n\pi}{3} \right)\frac{\partial}{\partial u_n} -\cos\left(\frac{(n+1)\pi}{3} \right) \frac{\partial}{\partial u_{n+1}}\right],\\
X_3&=&(-1)^n\left[\sin\left(\frac{n\pi}{3} \right) \frac{\partial}{\partial u_n} -\sin\left(\frac{(n+1)\pi}{3} \right) \frac{\partial}{\partial u_{n+1}}\right],\\
X_4&=&\frac{\partial}{\partial n}.
\end{eqnarray}
\textit{General solutions}:
If we assume that $v_n=v(n,u_n,u_{n+1})$ is an invariant of $X=Q\partial u_n + Q(n+1,u_{n+1})\partial u_{n+1}$, we may obtain the solution of (\ref{dp1=0}) by using the method of characteristics.
\begin{itemize}
\item Reduction using $X_1=u_n \frac{\partial}{\partial u_n} + u_{n+1} \frac{\partial}{\partial u_{n+1}}$.
Using the characteristic equation given by
\begin{eqnarray}
\frac{du_n}{u_n}=\frac{du_{n+1}}{{u_{n+1}}}=\frac{dv_n}{0},
\end{eqnarray}
one can readily check that $v_n = f\left(u_{n+1}/{u_{n}}\right)$
for some function $f$. We choose $f$ to be the identity function, i.e.,
\begin{eqnarray}\label{p1x1}
v_n = \frac{u_{n+1}}{{u_{n}}}
\end{eqnarray}
and we act the shift operator on $v_n$ to get the first-order difference equation
\begin{eqnarray}
v_{n+1} =-\frac{1}{{v_{n}}}-1
\end{eqnarray}
whose solution is given by
\begin{eqnarray}\label{p1x1a}
v_n& =& c_1e^{\frac{2\pi i}{3}}+ c_2  e^{\frac{-2\pi i}{3}}.
\end{eqnarray}
Equating (\ref{p1x1}) and (\ref{p1x1a}) we obtain
\begin{eqnarray}\label{p1x1aa}
u_{n+1}& =& e^{\frac{2\pi i}{3}}u_n \qquad \textit{and}\qquad  u_{n+1} = e^{\frac{-2\pi i}{3}}u_n.
\end{eqnarray}
We note that equation (\ref{dp1=0}) has been reduced by one order to (\ref{p1x1aa}) which follow a geometric progression. Therefore
\begin{eqnarray}
u_{n}& =& e^{\frac{2n\pi i}{3}}u_0 \qquad \textit{and}\qquad  u_n = e^{\frac{-2n\pi i}{3}}u_0
\end{eqnarray}
are solutions to (\ref{dp1=0}).
\item Reduction using $X_2 + i X_3$. 
    Applying the same methods we have  obtained
    \begin{eqnarray}
v_{n+1}& =& -e^{\frac{\pi i}{3}}v_n,
\end{eqnarray}
where
\begin{eqnarray}\label{x2x3}
v_n= u_{n+1}-e^{\frac{2i\pi}{3}}u_n.
\end{eqnarray}
Equation (\ref{x2x3}) is further reduced to
\begin{eqnarray}\label{p1sol2}
u_n=(-1)^{\frac{2(n-1)}{3}} \left(u_1-e^{\frac{2i\pi}{3}}u_0\right)\left[1+\frac{e^{\frac{-4i n\pi}{3}}(1-e^{\frac{4i n\pi}{3}})}{-1+(-1)^{\frac{2}{3}}} \right]
\end{eqnarray}
which is also a solution to equation (\ref{dp1=0}).
\item Naturally, the reduction using $X_2-iX_3$ leads to the conjugate of (\ref{p1sol2}).
\end{itemize}

\subsection{The discrete Painlev\'e II}
Consider the Painlev\'e equation II
\begin{equation}\label{dp2}
u_{n+2}=\omega = -u_n+\frac{u_{n+1}(an+b)+c}{1-{u_{n+1}}^2}.
\end{equation}
Finding the symmetry generator of (\ref{dp2})  requires us to determine the \textit{characteristic} of the one-parameter group, $Q$. Again, we assume that $Q$ depends on $n$ and $u_n$ only, and that the symmetry is of the form $X=\xi(n,u_n)\partial n+Q(n,u_n)\partial{u_n}+Q(n+1,u_{n+1})\partial{u_{n+1}}$. Here, the symmetry condition (\ref{symcdt}) becomes
\begin{eqnarray}\label{scp2}
&&Q(n+2,\omega)-\left[ \frac{(an+b)(1+{u_{n+1}}^2)+2cu_{n+1}}{(1-{u_{n+1}}^2)^2}\right]Q(n+1, u_{n+1})+Q(n,u_n)\nonumber\\&&-\frac{au_{n+1}}{1-{u_{n+1}}^2} \xi(n,u_n)= 0.
\end{eqnarray}
We now apply the \textit{differential operator}, $L$, given by
\begin{equation}\label{5.1.1G}
L=\partial u_n+\frac{\partial u_{n+1}}{\partial u_n}\partial u_{n+1},
\end{equation}
to equation (\ref{scp2}) and we differentiate the resulting equation with respect to $u_n$, keeping $u_{n+1}$ fixed, to get
\begin{eqnarray}\label{zx}
&& an Q'' -a\xi ' +bQ'' +  \left[(a^2n^2+2abn+b^2+2c)Q''-(a^2n+ab)\xi ''\right]u_{n+1}  \nonumber\\
&&- \left[ (an+b)Q''+2ac\xi''\right]{u_{k+1}}^2  -\big[ (a^2n^2+2abn+b^2+2c)Q''\nonumber\\
&&+(a^2n+ab)\xi \big]{u_{k+1}}^3+a\xi'{u_{k+1}}^4.
\end{eqnarray}
Similarly, after equating the coefficients of powers of $u_{n+1}$ to zero, we have solved the resulting determining system and we have noticed that condition for having non-zero characteristics that depend on $n$ and $u_n$ only is when $a=b=c=0$, that is, \begin{equation}\label{dp2=0}
u_{n+2}= -u_n.
\end{equation}
Under this condition, the symmetries are given by
\begin{eqnarray}
&& X_1=\partial _n,\qquad X_2=\alpha(n,u_n)\partial_{u_n}+\alpha(n+1,u_{n+1})\partial_{u_{n+1}},
\end{eqnarray}
where $\alpha$ satisfies the original equation (\ref{dp2=0}), i.e.,
\begin{equation}
\alpha(n+2,u_{n+2})= -\alpha(n,u_n).
\end{equation}
\textbf{Note}. If we assume that the function $\alpha$ does not depend on $u_n$, we obtain three symmetries:
\begin{eqnarray}\label{alphan}
&& X_1=\partial _n,\,X_2=\cos\left(\frac{n\pi}{2} \right) \frac{\partial}{\partial u_n} -\sin\left(\frac{n\pi}{2} \right) \frac{\partial}{\partial u_{n+1}},\nonumber\\
&&X_3=\sin\left(\frac{n\pi}{2} \right) \frac{\partial}{\partial u_n} +\cos\left(\frac{n\pi}{2} \right) \frac{\partial}{\partial u_{n+1}}.
\end{eqnarray}
\textit{General solutions}:
 We suppose that $v_n=v(n,u_n,u_{n+1})$ is an invariant of  (\ref{alphan}).
\begin{itemize}
\item Reduction using $X_2+i\,X_3$.\\
The characteristic equation
\begin{eqnarray}
\frac{du_n}{i^n}=\frac{du_{n+1}}{{i^{n+1}}}=\frac{dv_n}{0}
\end{eqnarray}
can be solved to get the invariants $\gamma = v_n$ and  $\alpha = u_{n+1}-e^{\left(\frac{i\pi}{2}\right)} u_n$.
We have,
\begin{eqnarray}\label{p2x2}
v_n =u_{n+1}-e^{\left(\frac{i\pi}{2}\right)} u_n
\end{eqnarray}
so
\begin{eqnarray}\label{p2x2a}
v_{n+1} = S(v_n)=e^{\left(\frac{i\pi}{2}\right)}v_n=e^{\frac{i(n+1)\pi}{2}}v_0.
\end{eqnarray}
Equations (\ref{p2x2}) and (\ref{p2x2a}) lead to a first-order difference equation
\begin{eqnarray}\label{p2x2b}
u_{n+1}& =&e^{\left(\frac{i\pi}{2}\right)} u_n+e^{\frac{i{n}\pi}{2}}\left( u_{1}-e^{\left(\frac{i\pi}{2}\right)} u_0\right).
\end{eqnarray}
It has to be noted that equation (\ref{dp2=0}) has been reduced by one order to (\ref{p2x2b}). We deduce that the general solution in this case is given by
\begin{eqnarray}\label{p1sol}
u_n= \left(u_1-i u_0\right)\left[\frac{3}{2}-\frac{1}{2} (-1)^n \right]i^{n-1}.
\end{eqnarray}
\item The reduction using $X_2-i\,X_3$ leads to the conjugate of (\ref{p1sol}).
\end{itemize}
\subsection{The discrete Painlev\'e III}
Consider the discrete painlev\'e equation III
\begin{equation}\label{dpe3}
u_{n+2}=\frac{1}{u_n}\left(\frac{au_{n+1}^2+bu_{n+1}+c}{u_{n+1}^2+du_{n+1}+e}\right),
\end{equation}
 where $a,\,b,\,c,\,d$ and $e$ are constant. We seek characteristics of the form $Q=Q(n,u_n)$. The symmetry condition (\ref{symcdt}) gives
\begin{eqnarray}\label{scdp3a}
&&Q(n+2,\omega)+\frac{au_{n+1}^2+bu_{n+1}+c}{u_n^2({u_{n+1}}^2+du_{n+1}+e)}Q(n, u_{n})\nonumber\\
&&-\frac{(ad-b)u_{n+1}^2+(2ae-2c)u_{n+1}+eb-cd}{u_n({u_{n+1}}^2+du_{n+1}+e)^2} Q(n+1, u_{n+1})= 0.
\end{eqnarray}
There are three separate pairs of arguments associated with the characteristic $Q$. 
Here, the procedure for obtaining the system of determining equations can be summarized as follows
\begin{itemize}
\item
Apply the differential operator $L$ on (\ref{scdp3a}).
\item
Multiply through to clear fractions.
\item
Differentiate with respect to $u_n$ twice, keeping $u_{n+1}$ fixed.
\item
Divide by ${u_{n+1}}^3$
\item
Separate by powers of $u_{n+1}$ and equate to zero.
\end{itemize}
These steps lead to the system
\begin{eqnarray}
{u_{n+1}}^5&:&(-2a^2d+2ab)Q''(n,u_n)=0,\\
{u_{n+1}}^4&:&(-4a^2d^2+2abd-4a^2e+2b^2+4ac)Q''(n,u_n)=0,\\
{u_{n+1}}^3&:&(a^2d-ab)u_nQ^{(3)}(n,u_n)+(-2a^2d^3-2abd^2-12a^2de+3a^2d\nonumber \\
&&+4b^2d+8acd-2abe-3ab+6bc)Q''(n,u_n)=0,
\end{eqnarray}\begin{eqnarray}
{u_{n+1}}^2&:&(a^2d^2+2a^2e-b^2-2ac)u_nQ^{(3)}(n,u_n)+(-2abd^3-8a^2d^2e\nonumber \\
&&+3a^2d^2+2b^2d^2+4acd^2d-12abde-8a^2e^2+14bcd+6a^2e+2b^2e\nonumber\\
&&+4ace-3b^2-6ac+4c^2)Q''(n,u_n)=0,\\
{u_{n+1}}&:&(abd^2+3a^2de-b^2d-2acd+2abe-3bc)u_nQ^{(3)}(n,u_n)+\nonumber \\
&&(-10abd^2e-10a^2de^2+3abd^2+10bcd^2+9a^2de-10abe^2-3b^2d\nonumber\\
&&-6acd+10c^2d+6abe+10bce-9bc)Q''(n,u_n)=0,\\
1&:&(4abde+2a^2e^2-4bcd-2c^2)u_nQ^{(3)}(n,u_n)+(2bcd^3-2b^2d^2e\nonumber \\
&&-4acd^2e-14abde^2-4a^2e^3+8c^2d^2+12abde+12bcde+6a^2e^2\nonumber\\
&&-2b^2e^2-4ace^2-12bcd+8c^2e-6c^2)Q(n,u_n)=0
\end{eqnarray}
which reduces to two possibilities:
\begin{eqnarray}
 ad=b,\quad ae=c
\end{eqnarray}
and
\begin{eqnarray}
Q''(n,u_n)=0.
\end{eqnarray}

\subsubsection{Case $ad=b$, $ae=c$}
Equation (\ref{dpe3}) simplifies to
\begin{equation}\label{p3x2s}
u_{n+2}=\frac{a}{u_n}
\end{equation}
and its symmetries are given by
\begin{eqnarray}
&&X_1=\left( 1-\frac{u_n^2}{a}\right) \frac{\partial}{\partial u_n} -\left( 1-\frac{u_{n+1}^2}{a}\right) \frac{\partial}{\partial u_{n+1}},\label{sympIII'1}\\
&&X_2=\sin\left(\frac{n\pi}{2} \right)\left( 1+\frac{u_n^2}{a}\right) \frac{\partial}{\partial u_n} +\cos\left(\frac{n\pi}{2} \right)\left( 1+\frac{u_{n+1}^2}{a}\right) \frac{\partial}{\partial u_{n+1}},\label{sympIII'2}\\
&&X_3=\cos\left(\frac{n\pi}{2} \right)\left( 1+\frac{u_n^2}{a}\right) \frac{\partial}{\partial u_n} -\sin\left(\frac{n\pi}{2} \right)\left( 1+\frac{u_{n+1}^2}{a}\right) \frac{\partial}{\partial u_{n+1}},\label{sympIII'3}\\
&&X_4=(-1)^n\left( 1-\frac{u_n^2}{a}\right) \frac{\partial}{\partial u_n} -(-1)^n\left( 1-\frac{u_{n+1}^2}{a}\right) \frac{\partial}{\partial u_{n+1}},\label{sympIII'4}\\
&&X_5=\cos\left(\frac{n\pi}{2} \right)u_n \frac{\partial}{\partial u_n} -\sin\left(\frac{n\pi}{2} \right)u_{n+1} \frac{\partial}{\partial u_{n+1}},\label{sympIII'5}\\
&&X_6=\sin\left(\frac{n\pi}{2} \right)u_n \frac{\partial}{\partial u_n} +\cos\left(\frac{n\pi}{2} \right)u_{n+1} \frac{\partial}{\partial u_{n+1}},\,X_7 =\frac{\partial}{\partial n}.\label{sympIII'6}
\end{eqnarray}
\textbf{Note.} By making the change of variable $w_n=\ln u_n$, equation (\ref{p3x2s}) reduces to (\ref{dp2=0}).
\subsubsection{Case $Q''(n,u_n)=0$}
The general solution in this case is given by
\begin{eqnarray}\label{cq''}
Q(n,u_n)=\alpha(n)u_n+\beta(n),
\end{eqnarray}
where $\alpha$ and $\beta$ are functions of $n$. The substitution of $Q$, given by  (\ref{cq''}), in equation (\ref{scdp3a}) leads to two cases:\newline
\textbf{Case $\beta=0$:} The characteristic becomes $Q=\alpha(n)u_n$ and for the equation (\ref{scdp3a}) to be satisfied we must have
\begin{eqnarray}
\alpha(n+2)=-\alpha(n),\quad ae=c,\quad ad=b.
\end{eqnarray}
Thus, equation (\ref{dpe3}) simplifies to $u_{n+2}=a/u_n$ which has two symmetries given by (\ref{sympIII'5}) and (\ref{sympIII'6}).\\
\textbf{Case $d^2=-2e$:} For the equation (\ref{scdp3a}) to be satisfied we must have
\begin{eqnarray}
\alpha(n+2)+\alpha(n+1)+\alpha(n)=0,\quad a=c=d=e=\beta=0.
\end{eqnarray}
Thus, the equation (\ref{dpe3}) simplifies to
\begin{equation}\label{p3x3t}
u_{n+2}=\frac{b}{u_nu_{n+1}}
\end{equation}
and the symmetries in this case are given by
\begin{eqnarray}
X_1&=&(-1)^n\left[\cos\left(\frac{n\pi}{3} \right)u_n \frac{\partial}{\partial u_n} -\cos\left(\frac{(n+1)\pi}{3} \right)u_{n+1} \frac{\partial}{\partial u_{n+1}}\right],\\
X_2&=&(-1)^n\left[\sin\left(\frac{n\pi}{3} \right)u_n \frac{\partial}{\partial u_n} -\sin\left(\frac{(n+1)\pi}{3} \right)u_{n+1} \frac{\partial}{\partial u_{n+1}}\right]\\
X_3&=& \frac{\partial}{\partial n}.
\end{eqnarray}
\textbf{Note.} By making the change of variable $U_n=\ln u_n$, equation (\ref{p3x3t}) reduces to (\ref{dp1'}) when $a$ and $b$ are equal to zero.
\subsection{The discrete Painlev\'e IV}
For Painlev\'e IV we shall consider the result obtained in paper \cite{Ramani} and we shall assume that $\gamma _0 = a=b=0$. Therefore, we may, without loss of generality, let the equation be of the form
\begin{eqnarray}\label{p4}
u_{n+2}=\omega =\frac{1}{u_n+u_{n+1}}\left( -u_n u_{n+1}+\frac{\mu}{{u_{n+1}}^2}+\epsilon_0\right).
\end{eqnarray}
Here, we present the results without details since the method is similar to the one presented above. After a set of long calculations we obtain the following symmetries
\begin{eqnarray}
&X_1=\cos\left(\frac{2n\pi}{3} \right)\left({u_n}^2+\epsilon_0\right) \frac{\partial}{\partial u_n} +\cos\left(\frac{2(n+1)\pi}{3} \right)\left({u_{n+1}}^2+\epsilon_0\right) \frac{\partial}{\partial u_{n+1}},\\
&X_2=\sin\left(\frac{2n\pi}{3} \right)\left({u_n}^2+\epsilon_0\right) \frac{\partial}{\partial u_n} +\sin\left(\frac{2(n+1)\pi}{3} \right)\left({u_{n+1}}^2+\epsilon_0\right) \frac{\partial}{\partial u_{n+1}},\\
&X_3=\frac{\partial}{\partial n}.\hspace{9.7cm}
\end{eqnarray}
\textbf{Note.} If we assume that $\mu$ and $\epsilon _0$ are zero, equation (\ref{p4}) becomes
\begin{eqnarray}\label{p4'}
u_{n+2} =-\frac{u_n u_{n+1}}{u_n+u_{n+1}}
\end{eqnarray}
and has three symmetries given as follows
\begin{eqnarray}
X_1 &=& u_n \frac{\partial}{\partial u_n}+ u_{n+1} \frac{\partial}{\partial u_{n+1}},\label{symp401}\\
X_2&=&(-1)^n\left[\cos\left(\frac{n\pi}{3} \right){u_n}^2 \frac{\partial}{\partial u_n} -\cos\left(\frac{(n+1)\pi}{3} \right){u_{n+1}}^2 \frac{\partial}{\partial u_{n+1}}\right],\\
X_3&=&(-1)^n\left[\sin\left(\frac{n\pi}{3} \right){u_n}^2 \frac{\partial}{\partial u_n} -\sin\left(\frac{(n+1)\pi}{3} \right){u_{n+1}}^2 \frac{\partial}{\partial u_{n+1}}\right],\\
X_4&=&\frac{\partial}{\partial n}.
\end{eqnarray}
The general solution of equation (\ref{p4'}) can be constructed by applying the method of characteristics using the symmetry $X_1$ given by (\ref{symp401}). In fact, the invariant $v_n=u_{n+1}/u_n$ obtained by using $X_1= u_n \partial/\partial u_n+ u_{n+1} \partial/\partial u_{n+1}$ satisfies the equation
\begin{eqnarray}\label{p4''}
v_n =\frac{u_{n+1}}{u_n}=-\frac{2}{2(-1)^nv_0-(-1)^n+1}.
\end{eqnarray}
The solution to the first-order difference equation (\ref{p4''}) is also the general solution to equation (\ref{p4'}). It is given by
\begin{eqnarray}\label{p4'''}
u_n =(-1)^{n-1}2^{n+\lceil \frac{1-n}{2} \rceil  -1}\left(2-2\frac{u_0}{u_1}\right)^{\lceil \frac{2-n}{2} \rceil-1}{\left(\frac{u_0}{u_1}\right)}^{\lceil \frac{1-n}{2} \rceil+1},
\end{eqnarray}
where $\lceil x \rceil $ is the ceiling function.
\subsection{The discrete Painlev\'e V}
The discretization of the continuous case of the Painlev\'e V was studied in \cite{Ramani} and the discrete form was given. Our aim is to find the symmetries of the discrete form. To ease the computation we shall assume that the parameters $\alpha_0,\, \sigma,\, \rho_0,\, \theta$ and $\mu$ are equal to zero, that is, we consider the equation
\begin{eqnarray}\label{P5eq}
u_{n+2}=\omega =\frac{u_n u_{n+1}}{2u_{n+1}u_n-u_{n}-u_{n+1}}.
\end{eqnarray}
If we impose the invariance condition, assuming that the characteristic is of the form $Q=Q(n,u_n)$, we obtain
\begin{eqnarray}\label{P5a}
Q(n+2, \omega)+ \frac{1 }{\left(2u_{n+1}u_n-u_{n}-u_{n+1}\right)^2}\left[ {u_{n+1}}^2 Q+{u_n}^2Q(n+1,u_{n+1})\right].
\end{eqnarray}
Again we proceed by applying the operator $L$ to equation (\ref{P5a}) and then we differentiate with respect to $u_n$ four times to get a system differential equations that involves $u_n$ and a function of $u_n$ only. The lengthy calculations for solving the resulting system lead to the following symmetries:
\begin{eqnarray}
X_1 &=& \left(u_n-\frac{2}{3}{u_n}^2 \right)\frac{\partial}{\partial u_n}+ \left(u_{n+1}-\frac{2}{3}{u_{n+1}}^2 \right) \frac{\partial}{\partial u_{n+1}},\\
X_2&=&(-1)^n\left[\cos\left(\frac{n\pi}{3} \right){u_n}^2 \frac{\partial}{\partial u_n} -\cos\left(\frac{(n+1)\pi}{3} \right){u_{n+1}}^2 \frac{\partial}{\partial u_{n+1}}\right],\\
X_3&=&(-1)^n\left[\sin\left(\frac{n\pi}{3} \right){u_n}^2 \frac{\partial}{\partial u_n} -\sin\left(\frac{(n+1)\pi}{3} \right){u_{n+1}}^2 \frac{\partial}{\partial u_{n+1}}\right],\\
X_4&=&\frac{\partial}{\partial n}.
\end{eqnarray}
\section{Conclusion}
\noindent We have presented a procedure for obtaining symmetries of difference equations. We have assumed that $\xi$ and $Q$ are functions of $n$ and $u_n$ only (not all equations admit such characteristics) and we have derived a number of symmetries that were used, in some cases, to solve the difference equations. In the case of Painlev\'e I we have  shown that $\xi$ is not zero whenever $a$ is non-zero. 


\begin{thebibliography}{22}
\bibitem{Hydon1} P. E. Hydon. Symmetries and first integrals of ordinary difference equations. \textit{ Proc. Roy. Soc. Lond. A } \textbf{456} (2000), 2835-2855.

    \bibitem{FK2}M. Folly-Gbetoula, L. Ndlovu, A.H. Kara and A. Love.
Symmetries, Associated First Integrals, and Double Reduction of Difference Equations. \textit{Abstract and Applied Analysis} \textbf{2014}, Article ID 490165.
\bibitem{s} G. R. W. Quispel and R. Sahadevan. Lie symmetries and the integration of difference equations.\textit{ Physics Letters A} \textbf{184} (1993), 64-70.


\bibitem{Ramani} A. Ramani. Discrete Versions of the Painlev\'e Equations. \textit{Physical letter} \textbf{14},1991.
\end{thebibliography}
\end{document}